\documentclass{article}

\usepackage{arxiv}

\usepackage[utf8]{inputenc} % allow utf-8 input
\usepackage[T1]{fontenc}    % use 8-bit T1 fonts
\usepackage{hyperref}       % hyperlinks
\usepackage{url}            % simple URL typesetting
\usepackage{booktabs}       % professional-quality tables
\usepackage{amsfonts}       % blackboard math symbols
\usepackage{nicefrac}       % compact symbols for 1/2, etc.
\usepackage{microtype}      % microtypography
\usepackage{lipsum}		% Can be removed after putting your text content
\usepackage{graphicx}
\usepackage{natbib}
\usepackage{doi}
\usepackage{amsmath}
\usepackage{bm}

\title{Model Selection for Exposure-Mediator Interaction \\ \large{For the Alzheimer’s Disease Neuroimaging Initiative} \thanks{Data used in the preparation of this article were obtained from the Alzheimer's Disease Neuroimaging Initiative (ADNI) database (adni.loni.usc.edu). As such, the investigators within the ADNI contributed to the design and implementation of ADNI and/or provided data but did not participate in the analysis or writing of this report. A complete listing of ADNI investigators can be found at: \url{http://adni.loni.usc.edu/wp-content/uploads/how_to_apply/ADNI_Acknowledgement_List.pdf}}
}

\date{} 					% Or removing it

\author{ \href{https://orcid.org/0000-0002-8635-2669}{\includegraphics[scale=0.06]{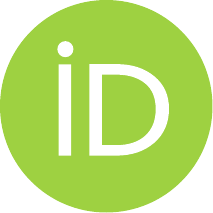}\hspace{1mm}Ruiyang Li} \\
	Department of Biostatistics\\
	Columbia University\\
	New York, NY \\
	\texttt{rl3034@cumc.columbia.edu} \\
	\And
	Xi Zhu \\
	Department of Psychiatry\\
	Columbia University, New York State Psychiatric Institute\\
	New York, NY \\
	\texttt{xi.zhu@nyspi.columbia.edu} \\
	\And
	\href{https://orcid.org/0000-0003-3177-6357}{\includegraphics[scale=0.06]{orcid.pdf}\hspace{1mm}Seonjoo Lee} \thanks{To whom correspondence should be addressed} \\
	Department of Biostatistics and Psychiatry\\
	Columbia University, New York State Psychiatric Institute\\
	New York, NY \\
	\texttt{seonjoo.lee@nyspi.columbia.edu} \\
}

\renewcommand{\shorttitle}{Model Selection for Exposure-Mediator Interaction}

\hypersetup{
pdftitle={Model Selection for Exposure-Mediator Interaction},
pdfsubject={stat.ME, stat.CO, stat.AP},
pdfauthor={Ruiyang Li, Xi Zhu, Seonjoo Lee},
pdfkeywords={Brain compensation; Exposure by mediator interaction; Hierarchical structure; High-dimensional mediators; Model selection; Mediation analysis; Neuroimaging},
}

\begin{document}
\maketitle

\begin{abstract}
In mediation analysis, the exposure often influences the mediating effect, i.e., there is an interaction between exposure and mediator on the dependent variable. When the mediator is high-dimensional, it is necessary to identify non-zero mediators (M) and exposure-by-mediator (X-by-M) interactions. Although several high-dimensional mediation methods can naturally handle X-by-M interactions, research is scarce in preserving the underlying hierarchical structure between the main effects and the interactions. To fill the knowledge gap, we develop the XMInt procedure to select M and X-by-M interactions in the high-dimensional mediators setting while preserving the hierarchical structure. Our proposed method employs a sequential regularization-based forward-selection approach to identify the mediators and their hierarchically preserved interaction with exposure. Our numerical experiments showed promising selection results. Further, we applied our method to ADNI morphological data and examined the role of cortical thickness and subcortical volumes on the effect of amyloid-beta accumulation on cognitive performance, which could be helpful in understanding the brain compensation mechanism.
\end{abstract}

% keywords can be removed

\section{Introduction}\label{sec1}

\begin{figure}[h]
\centering\includegraphics[scale=0.35]{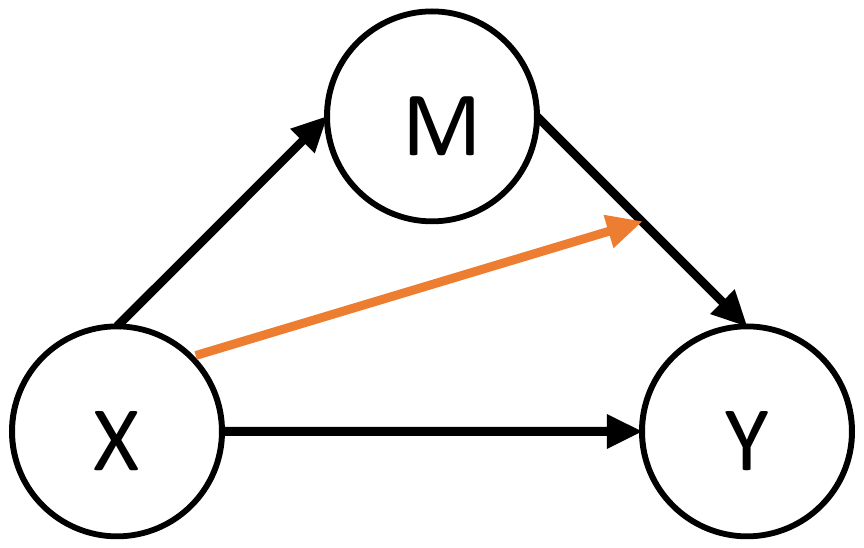}
\caption{A graphical representation of the mediation effect of the intervening variable M on the relationship between the independent variable X and the dependent variable Y (black arrows, upper triangular part) and the interaction effect between X and M (orange arrow)}
\label{Fig1}
\end{figure}

A mediation model examines how an independent variable or an exposure (X) affects a dependent variable (Y) through one or more intervening variables or mediators (M) \citep{baron_moderatormediator_1986,robins_identifiability_1992,holland_causal_1988,mackinnon_introduction_2008,sobel_identification_2008,preacher_asymptotic_2008,imai_identification_2010,vanderweele_controlled_2011,imai_identification_2013,pearl_direct_2013,vanderweele_mediation_2014,vanderweele_explanation_2015,daniel_causal_2015}. In the mediation mechanism, we often observe that X influences the mediating effect of M on Y, i.e., there is an interaction between X and M on the dependent variable (Y), as represented in Figure \ref{Fig1}. 

In the single mediator case, \cite{vanderweele_unification_2014} proposed the 4-way decomposition method to handle the X-by-M interaction, which was used in \cite{wang_association_2019} to find the mediated interaction effect. This decomposition idea was later extended to the few mediators case, such as in \cite{vanderweele_mediation_2014} and \cite{bellavia_decomposition_2018}.

New mediation methods have also been developed to address the issue of high dimensions in the mediators. Some used regularization-based methods  \citep{zhao_pathway_2016,serang_exploratory_2017,li_regularized_2021}. Others used hybrid methods such as the combined filter method with coordinate descent algorithm \citep{van_kesteren_exploratory_2019}, screening and regularization \citep{zhang_estimating_2016,zhang_mediation_2021,luo_high-dimensional_2020,schaid_penalized_2020}, and dimension reduction and regularization  \citep{zhao_sparse_2020}.

For models involving interactions with high-dimensional data, it is important to preserve the underlying hierarchical structure between the main effects and the interactions \citep{hao_note_2017,nelder_reformulation_1977}.     
In regression settings, regularization methods with different penalty functions \citep{efron_least_2004,yuan_structured_2009,zhao_composite_2009,choi_variable_2010,bien_lasso_2013} have been developed to aid the interaction selection, but they become infeasible as the number of independent variables increases \citep{hao_model_2018}. To address such limitation, \cite{hao_model_2018} proposed an efficient interaction selection method for high-dimensional data, the regularization algorithm under marginality principle (RAMP), in which possible interaction terms are sequentially added based on the current main effects. 

Compared to the regression settings, it is more challenging to preserve the hierarchical structure between the main effects and interactions in the mediation analysis because model selection now involves two models: M on X and Y on M. Although the high-dimensional mediation methods mentioned earlier may naturally handle X-by-M interactions, research is scarce in preserving such hierarchy. Therefore, in this paper, we aim to identify the mediators and the exposure by mediator interactions in the high-dimensional mediators setting while addressing the underlying hierarchical structure between the main effects and interactions. 

This identification of mediators and their hierarchically preserved interaction with exposure can be useful in explaining how the brain reacts during the process of brain-related changes and in understanding the brain compensation mechanism, defined as the phenomenon that the effect of the brain on the outcome is altered. 
For example, under the context of cognitive aging and brain pathology, where we have cognitive or clinical performance as the outcome, aging or pathology as the exposure, and certain brain measures such as cortical thickness as the potential mediators, we often observe that thinner cortical thickness (M) is associated with the worse cognitive performance (Y), and further, in the presence of some pathology (X), we observe the opposite association or no association anymore. In other words, the brain acts differently on the outcome because of the exposure (i.e., there is X-by-M interaction), which illustrates the brain compensation mechanism based on our definition. Or alternatively, we may think that the brain tries to compensate for the loss in cognition due to the pathology, which is also indicative of the idea of compensation. 

Our proposed method employs a sequential regularization-based forward-selection approach to identify the mediators and their interactions with exposure while preserving the hierarchical structure between them. We will sequentially update the selected mediator and interaction set (initially empty) based on the selection in the previous step. For example, if a mediator is identified but its corresponding interaction term is not, we will include the identified mediator into the selected mediator set in the next step while keeping the selected interaction set as it is; if both the mediator and its corresponding interaction are identified, we will include them into the selected mediator set and the selected interaction set, respectively; if an interaction is identified, but its involving M is not, we will not only include the interaction into the selected interaction set but also include the involving M into the mediator set to preserve the hierarchical structure. Recently, \cite{wang_estimating_2020} proposed a method designed specifically for high-dimensional compositional microbiome mediators, in which the selection of the interaction between treatment and mediators was considered. In contrast to their method, our method focuses on the continuous mediators, which do not have any sum constraint on the mediators as in the microbiome data. In addition, instead of adding overall penalties to the objective function to preserve the hierarchical structure as in  \cite{wang_estimating_2020}, we adopt an adaptive way to update the penalties and include the potential mediator if either its interaction with the exposure is selected or its main effect is selected in the previous step. 

In this paper, we first introduced our proposed algorithm in Section \ref{sec2}. Then, we presented the simulation results in Section \ref{sec3}. Finally, we applied our method to the real-world human brain imaging data and examined the role of cortical thickness and subcortical volumes on the relationship between amyloid beta accumulation and cognitive performance in Section \ref{sec4}.

\section{Methods}
\label{sec2}

We aim to identify the mediators (M) and their interactions with exposure (X) while preserving the hierarchical structure between the main effects and interaction effects. The multivariate mediation model that we consider takes the following form. For each subject $i=1, ..., n$, 
\begin{gather*}
    \mathbf{M}_i = \boldsymbol{a_0} + \boldsymbol{a} X_i + {\boldsymbol{e_1}}_{i}, \\ 
    Y_i = c_0 + c X_i + \mathbf{M}_i^\top \boldsymbol{b_1} + (X_i \times \mathbf{M}_{i}^\top) \ \boldsymbol{b_2} + e_{2_i}, 
\end{gather*} 
where $\mathbf{M}_i = (M_{i,1} \ \cdots \ M_{i,V})^\top$, $\boldsymbol{a_0} = (a_{0_1} \ \cdots \ a_{0_V})^\top$, $\boldsymbol{a} = (a_{1} \ \cdots \ a_{V})^\top$, $\mathbf{e_1}_{i} = (e_{1_{i,1}} \ \cdots \ e_{1_{i,V}})^\top$ $\sim {MVN}(\mathbf{0}, \Sigma)$, $\boldsymbol{b_1} = (b_{1_1} \ \cdots \ b_{1_V})^\top$, $\boldsymbol{b_2}= (b_{2_1} \ \cdots \ b_{2_V})^\top$, and $e_{2_i} \sim N(0,\sigma^2)$. 

$X$ is the independent variable or the exposure of interest, $Y$ is the dependent variable or the outcome of interest, and $M_v$'s, where $v = 1, ..., V$, are the variables that might have the potential to mediate the effect of $X$ on $Y$. $X \times M_v$ interaction terms are included to investigate whether $X$ changes the way how $M_v$ affects $Y$. Any $M_v$ with non-zero $a_v$ and $b_{1_v}$ coefficients is considered to be a mediator. Any $X \times M_v$ with non-zero $b_{2_v}$ coefficient is considered to have an interaction effect. 

The joint distribution of the outcome and the mediator can be written as $f_{Y,M \mid \cdot} (y,m \mid \cdot) = f_{Y \mid M,\cdot} (y \mid m,\cdot) \ f_{M \mid \cdot}(m \mid \cdot).$ Under the Gaussian assumption of the error terms, the log-likelihood is given as 
\begin{eqnarray*}
\ell({\boldsymbol{\theta}}) 
& = & 
\sum_{i=1}^{n} \ell_{Y,\mathbf{M} \mid \boldsymbol{\theta}}(\boldsymbol{\theta} \mid Y_i,\mathbf{M}_i) 
  = 
  \sum_{i=1}^{n} \ell_{Y \mid \mathbf{M},\boldsymbol{\theta}}(\boldsymbol{\theta} \mid Y_i,\mathbf{M}_i) + \sum_{i=1}^{n} \ell_{ \mathbf{M} \mid \boldsymbol{\theta}}(\boldsymbol{\theta} \mid \mathbf{M}_i) \\
& = & 
  -n \log(2 \pi) - \frac{n}{2} \log \sigma^2 + \frac{n}{2} \log \mid \Omega \mid \\
  &&- \frac{1}{2\sigma^2} \sum_{i=1}^n \Big( Y_i -c_0 - cX_i - \mathbf{M}_i^\top \boldsymbol{b_1} - (X_i \times \mathbf{M}_{i}^\top) \ \boldsymbol{b_2} \Big)^2 \\
  &&- \frac{1}{2} \sum_{i=1}^n (\mathbf{M}_i - \boldsymbol{a_0} - \boldsymbol{a} X_i)^\top \Omega (\mathbf{M}_i - \boldsymbol{a_0} - \boldsymbol{a} X_i), 
\end{eqnarray*}
where $\Omega = \Sigma^{-1}$.

We denote $\mathcal{M} = \{1,2,...,V\}$ be the index set for all of the $V$ potential mediators $\mathbf{M}$, $\mathcal{M}_{k}$ be the index set for the mediators identified in step $k$, $\mathcal{M}_{k}^{c} = \mathcal{M}-\mathcal{M}_{k}$ be the index set for the remaining M variables (out of $V$ variables), $\mathcal{I}_{k}$ be the index set for the involving M variables in the interaction terms identified in step $k$, and $\mathcal{I}_{k}^{c} = \mathcal{M}-\mathcal{I}_{k}$ be the index set for the involving M variables in remaining interaction terms (out of $V$ variables). %We say that a $M_v$ variable is in the selected mediator set at step $k$ if its index $v$ is in $\mathcal{M}_{k}$ and that a $X \times M_v$ variable is in the selected interaction set at step $k$ if its index $v$ is in $\mathcal{I}_{k}$. 

Our algorithm is described as follows. Table \ref{Table1} provides a summary of the algorithm.

\textit{1. Initialization}

First, we standardize the data and compute $\lambda_{max} = n^{-1} \max \mid \mathbf{ [ X \ M \ X \times M ] }^\top \mathbf{Y} \mid$ and $\lambda_{min} = \zeta \lambda_{max}$ for some small $\zeta > 0$ (e.g., 0.05). Based on the determined $\lambda_{max}$ and $\lambda_{min}$, we generate an exponentially decaying sequence with length $K$ (e.g. 20), $\lambda_{max} = \lambda_{1} > \lambda_{2} > \cdots > \lambda_{K} = \lambda_{min}$. Also, we start with $\mathcal{M}_0 = \emptyset$ and $\mathcal{I}_0 = \emptyset$. 

\textit{2. Find regularization path}

For each $k = 1, ..., K$, we minimize the following objective function (\ref{obj_func}), in the form of negative log-likelihood plus penalties, with respect to $\boldsymbol{a,b_1,b_2}$. 
\begin{align}\label{obj_func}
    n \log \sigma^2 &- n \log \mid \Omega \mid \nonumber + \sum_{i=1}^n (\mathbf{M}_i - \boldsymbol{a_0} - \boldsymbol{a}X_i)^\top \Omega (\mathbf{M}_i - \boldsymbol{a_0} - \boldsymbol{a}X_i) \nonumber \\
&+ \frac{1}{\sigma^2} \sum_{i=1}^{n} \Big( Y_i - c_0 - c X_i - \mathbf{M}_i^\top \boldsymbol{b_1} - (X_i \times \mathbf{M}^\top_{i}) \boldsymbol{b_2} \Big)^2 \nonumber \\
&+ \lambda_k \mid \mid \boldsymbol{b}_{1_{v \in \mathcal{M}_{k-1}^{c} \cup \mathcal{I}_{k-1}^{c} } } \mid \mid_1 
+ \lambda_k \mid \mid \boldsymbol{b}_{2_{v \in \mathcal{I}_{k-1}^{c}} } \mid \mid _1 
+ \lambda_k \mid \mid \boldsymbol{a}_{v \in \mathcal{M}_{k-1}^{c} \cup \mathcal{I}_{k-1}^{c} } \mid \mid _1
\end{align} 
The penalization at current step $k$ is imposed based on the selected mediator set and the selected interaction set from the previous step (or, $\mathcal{M}_{k-1}$ and $\mathcal{I}_{k-1}$). If a $M_v$ variable is selected as a mediator in the previous step (i.e., the index $v$ is in $\mathcal{M}_{k-1}$), then we will not penalize the corresponding $a_{v}$ and $b_{1_v}$ at the current step. If a $X \times M_v$ variable is selected as an interaction in the previous step (i.e., the index $v$ is in $\mathcal{I}_{k-1}$), then we will not only not penalize the corresponding $b_{2_v}$ but also not penalize the corresponding $a_{v}$ and $b_{1_v}$ to force the main effect to be included into the model so that the hierarchical structure is preserved. These altogether form the penalty terms in the objection function (\ref{obj_func}). 

We utilize an iterative estimation approach to estimate the nuisance parameters $\sigma^2$ and $\Omega$ using QUIC R package and $\boldsymbol{a,b_1,b_2}$ using glmnet R package. We then update the selected mediator set and the selected interaction set correspondingly based on the estimated coefficients. The mediators identified (i.e. the $M_v$ variables with non-zero $a_v$ and $b_{1_v}$ coefficients) at current step $k$ form the current selected mediator set (i.e., their indices $v$'s form $\mathcal{M}_{k}$). The interactions identified (i.e. the $X \times M_v$ variables with non-zero $b_{2_v}$ coefficient) at current step $k$ form the current selected interaction set (i.e., their indices $v$'s form $\mathcal{I}_{k}$). 

\textit{3. Model selection}

Then, we compute Haughton's Bayesian information criterion (HBIC; \cite{haughton_choice_1988}) between the current model and the null model. HBIC for two model comparison can be computed as $HBIC = 2[ \ell({\boldsymbol{\hat{\theta_2}}}) - \ell({\boldsymbol{\hat{\theta_1}}}) ] - (d_2-d_1) \log \Big( \frac{N}{2\pi} \Big),$ where $\ell({\boldsymbol{\hat{\theta_j}}})$ is the log-likelihood function of the model $j$, $d_j$ is the number of parameters of the model $j$, and $N$ is the sample size \citep{bollen_bic_2014}. HBIC is chosen to evaluate the model performance instead of cross-validation because cross-validation may not perform well with a limited sample size. Previous studies have shown that HBIC stands out in the selection of measurement models \citep{haughton_information_1997,lin_selecting_2017}; among the information criterion (IC) measures, the scaled unit information prior BIC (SPBIC) and the HBIC have the best overall performance in choosing the true full structural models \citep{bollen_bic_2014,lin_selecting_2017}; SPBIC and HBIC performed the best in selecting path models and were recommended for model comparison in structural equation modeling (SEM) \citep{lin_selecting_2017}; HBIC might be preferable to SPBIC for its simplicity in computation \citep{lin_selecting_2017}. We select the model with the smallest HBIC as the final model.\\

Note that in the initialization step, $\lambda_{max}$ should be set to ensure that we start with the null model. In occasional cases, the computed $\lambda_{max}$ can be too small to give a null model as the starting point. To account for that, our algorithm gradually enlarges $\lambda_{max}$ by a factor (1.5 by default, which is a 50\% increase from the previous one) until we start with a null model.

\section{Simulation}
\label{sec3}

For each subject $i = 1, \cdots, N$, the exposure $X_i$ was independently generated from the standard normal distribution, each potential mediator was independently generated as $M_{i,v} = a_v X_i + \varepsilon_{1_{i,v}}$, where $\varepsilon_{1_{i,v}} \stackrel{i.i.d.}{\sim} \  N(0,1)$, and the outcome $Y_i$ was generated as $Y_i = 1 \cdot X_i + \Sigma_v b_{1_{v}} M_{i,v} + \Sigma_v b_{2_{v}} X_{i} \times M_{i,v} + \varepsilon_{2_{i}}$, where $\varepsilon_{2_{i}} \stackrel{i.i.d.}{\sim} \ N(0,1)$. We set the first three $M$ variables ($M_1, M_2, M_3$) to be the true mediators (i.e., having non-zero $a$ and $b_1$ coefficients) and set $X \times M_1$ to be the true exposure-by-mediator interaction term (i.e., having non-zero $b_2$ coefficient). 

In this simulation, the sample size was set to be $N = 100, 200, 400$, and the number of potential mediators was set to be $V = 50, 100, 200, 400$. The effect size (ES) represents the value of $a, b_1, b_2$ of the truth, which are $a_1, a_2, a_3, b_{1_{1}}, b_{1_{2}}, b_{1_{3}}, b_{2_{1}}$ in our case. We set $ES = 0.25, 0.5, 0.75, 1$ and set all other coefficient values to 0. Also, we used the default $K = 20$ and $\zeta = 0.05$ to generate the $\lambda$ sequence. 
The final model is given by the $\lambda$ that minimizes HBIC. 

Under each simulation scenario, we calculated the average true positive rate (TPR) and the average false discovery rate (FDR) across the 100 simulation runs for the mediator and the interaction, respectively. For each simulation run, the TPR was computed as the proportion of the truth that was selected by the algorithm. For example, the TPR for the mediator is $${\mbox{TPR}_{\mbox{med}}} = \frac{\mbox{the number of selected true mediators}}{\mbox{the number of true mediators}},$$ and the TPR for the interaction is $${\mbox{TPR}_{\mbox{int}}} = \frac{\mbox{the number of selected true interactions}}{\mbox{the number of true interactions}}.$$ The FDR was computed as the proportion of the falsely selected non-truth variables from the selection. That is, the FDR for the mediator is  $${\mbox{FDR}_{\mbox{med}}} = \frac{\mbox{the number of falsely selected mediators}}{\mbox{the number of selected mediators}},$$ and the FDR for the interaction is $${\mbox{FDR}_{\mbox{int}}} = \frac{\mbox{the number of falsely selected interactions}}{\mbox{the number of selected interactions}}.$$ 

\begin{figure}[h]
\centering\includegraphics[scale=0.65]{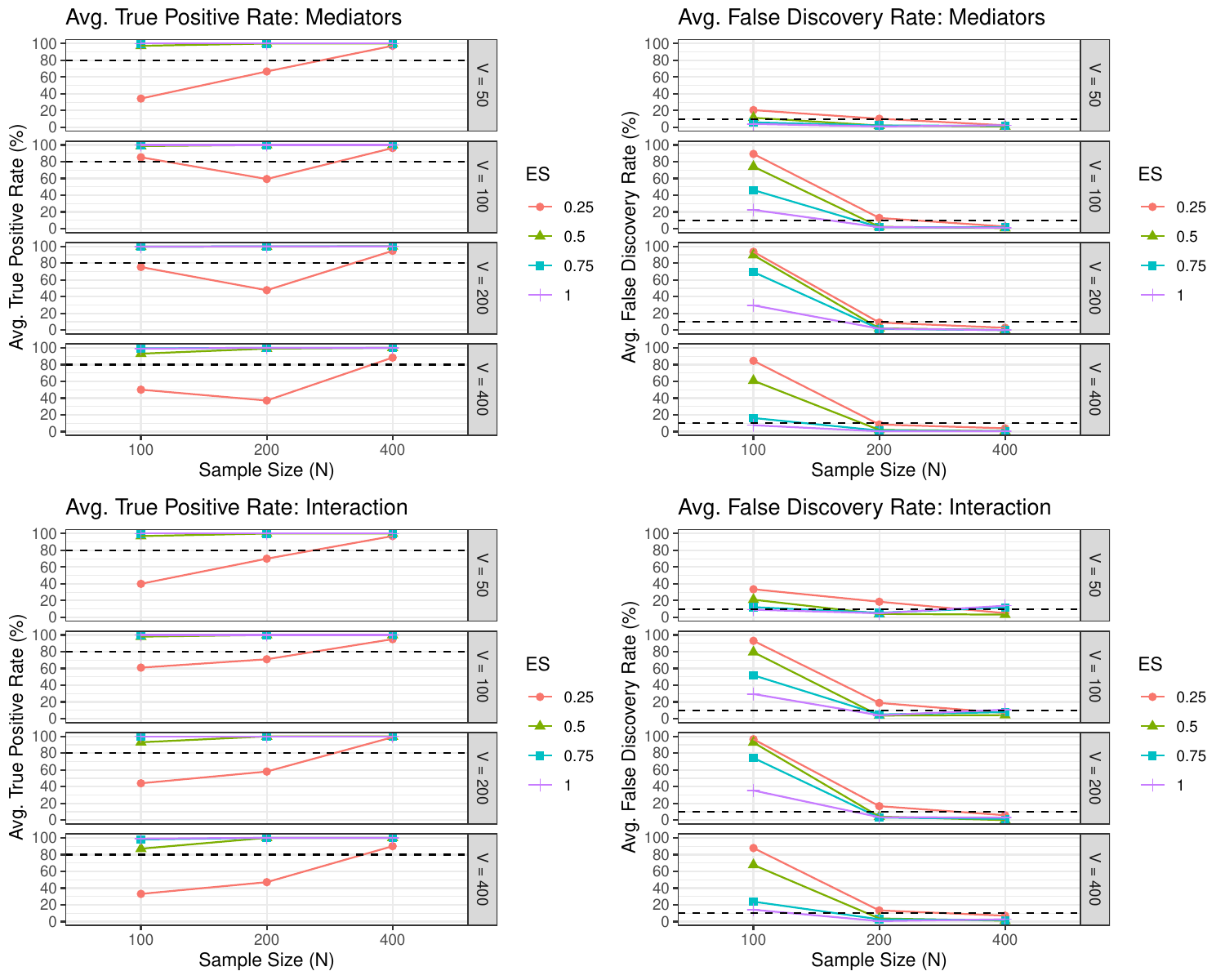}
\caption{The average True Positive Rate (TPR) and the average False Discovery Rate (FDR) across the 100 simulation runs by the sample size (N), the number of potential mediators (V) and the effect size (ES), for (a) the mediator (upper panel) and for (b) the interaction (lower panel), respectively}
\label{Fig2}
\end{figure}

Figure \ref{Fig2} shows the average TPR and the average FDR across the 100 simulation runs by the sample size, the number of potential mediators and the effect size, for (a) the mediator and for (b) the interaction, respectively. By the forward-selection feature of our algorithm, the hierarchical structure between interactions and mediators is preserved. Our results show that the average TPR generally increases with effect size, for both the mediator and the interaction. When the effect size and the sample size were moderate to large (N at least 200, ES at least 0.5), our algorithm can almost 100\% of the time identify all of the three true mediators and the true interaction term (TPR close to 100\%) and it is not likely to falsely select the non-truth (FDRs controlled under approximately 10\%). Also, we observed that by increasing the sample size, we may be able to make up for a small effect size to some degree and maintain a reasonably high TPR.

\section{Data Application}
\label{sec4}

We applied our algorithm to the human brain imaging data from the Alzheimer’s Disease Neuroimaging Initiative (ADNI), a longitudinal multicenter study designed for the early detection and tracking of Alzheimer’s disease. Specifically, we assessed the role of cortical thickness and subcortical volumes on the relationship between amyloid beta accumulation and cognitive abilities. We used the data from the participants without dementia -- diagnosed as mild cognitive impairment (MCI; N = 458) or cognitively normal (CN; N = 276). Participants' characteristics are displayed in Table \ref{Table2}.

The data were downloaded from the ADNI database (\url{http://adni.loni.usc.edu}). The initial phase (ADNI-1) recruited 800 participants, including approximately 200 healthy controls, 400 patients with late MCI, and 200 patients clinically diagnosed with AD over 50 sites across the United States and Canada and followed up at 6- to 12-month intervals for 2--3 years. ADNI has been followed by ADNI-GO and ADNI-2 for existing participants and enrolled additional individuals, including early MCI. To be classified as MCI in ADNI, a subject needed an inclusive Mini-Mental State Examination score of between 24 and 30, subjective memory complaint, objective evidence of impaired memory calculated by scores of the Wechsler Memory Scale Logical Memory II adjusted for education, a score of 0.5 on the Global Clinical Dementia Rating, absence of significant confounding conditions such as current major depression, normal or near-normal daily activities, and absence of clinical dementia.

All studies were approved by their respective institutional review boards and all subjects or their surrogates provided informed consent compliant with HIPAA regulations.

The exposure of interest is amyloid beta accumulation. CSF amyloid beta (A$\beta$-42) concentrations were measured in picograms per milliliter (pg/mL) by ADNI researchers using the highly automated Roche Elecsys immunoassays on the Cobas e601 automated system following extensive validation studies \citep{bittner2016technical, shaw2016method}. The CSF data used in this study were obtained from the ADNI files \verb |UPENNBIOMK9_04_19_17.csv|. Detailed descriptions of CSF acquisition including lumbar puncture procedures, measurement, and quality control procedures were presented in \url{http://adni.loni.usc.edu/methods/}.

The outcome variable of interest is the memory composite score. We used ADNI’s pre-generated cognitive composite scores that were constructed based on bi-factor confirmatory factor analyses models \citep{crane2012development}. Composite memory scores were derived using the Rey Auditory Verbal Learning Test, AD Assessment Schedule-Cognition, Mini-Mental State Examination, and Logical Memory.

MPRAGE T1-weighted MR images were used in this analysis. Cross-sectional image processing was performed using FreeSurfer Version 7.0.1. Region of interest (ROI)-specific cortical thickness and volume measures were extracted from the automated FreeSurfer anatomical parcellation using the Desikan-Killiany Atlas \citep{desikan2006automated} for cortical regions and ASEG \citep{fischlWholeBrainSegmentation2002} for subcortical regions. 
We considered 89 potential mediators that were derived from 68 cortical thickness measures from both left and right hemispheres, 5 corpus callosum subregion volume measures, and 16 subcortical volume measures from both left and right hemispheres (Thalamus, Caudate, Putamen, Pallidum, Hippocampus, Amygdala, Accumbens, VentralDC). Since the volume measures are typically confounded by brain size, we divided them by the estimated total intracranial volume to adjust for the potential confounding effect. Further, as all these 89 measures are often correlated with sex and age, we regressed them on sex and age and used the residuals as our potential mediators.

\begin{figure}[h]
\centering\includegraphics[scale=0.8]{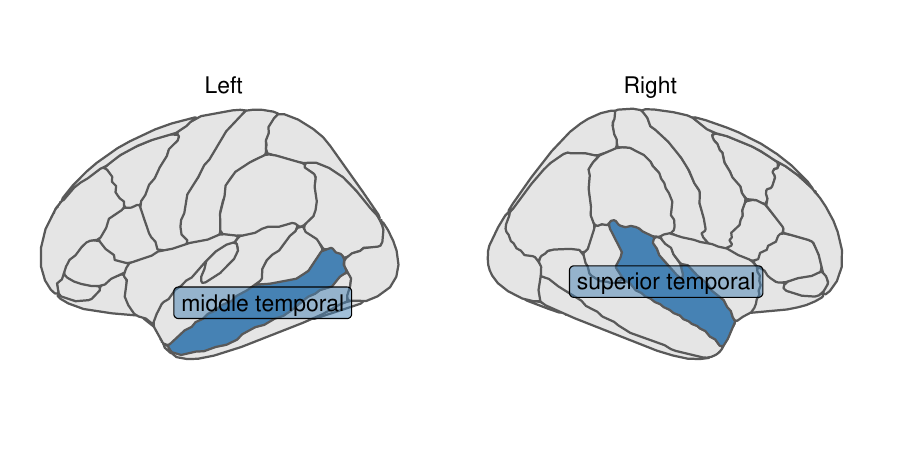}
\caption{Two cortical thickness measures (one in the middle temporal region in the left hemisphere and the other in the superior temporal region in the right hemisphere) that were identified to have the mediated interaction effects on the relationship between amyloid beta accumulation and memory using XMInt on ADNI dataset}
\label{Fig3}
\end{figure}

Our algorithm identified mediated interaction effects of cortical thickness from two temporal regions: the middle temporal region in the left hemisphere and the superior temporal region in the right hemisphere, as shown in Figure \ref{Fig3}. In other words, the cortical thickness of the two identified regions mediates the relationship between amyloid beta accumulation and cognitive abilities, and further, their effect on cognition is influenced by amyloid beta accumulation. 
The direction of the interaction can also be explored using the 4-way decomposition idea proposed in \cite{vanderweele_unification_2014} and \cite{vanderweele_mediation_2014}. Under the counterfactual independence assumptions, it showed that when there is no amyloid beta accumulation, the thinner cortex in the middle temporal region of the left hemisphere and the superior temporal region of the right hemisphere is associated with worse cognitive performance (PIE = 0.036, p < 0.01); however, when there is more amyloid beta accumulation involved, such mediation effect disappears -- the thinner cortex is no longer associated with worse cognition (TIE = 0.010, p = 0.1). These altogether illustrate the brain compensation mechanism during this process of brain-related changes.

\section{Discussion}
\label{sec5}

In this paper, we proposed the XMInt algorithm to identify the mediators and the exposure by mediator interactions in the high-dimensional mediators setting while preserving the underlying hierarchical structure between the main effects and the interaction effects, and we investigated the conditions when our algorithm demonstrates ideal model selection performance. Based on our simulation results, when the effect size and the sample size are moderate to large, our algorithm was able to correctly identify the true mediators and interaction almost all the time, without falsely selecting many non-truth variables.
To illustrate our algorithm, we also applied our method to real-world human brain imaging data. Two cortical thickness measures (one in the middle temporal region in the left hemisphere and the other in the superior temporal region in the right hemisphere) were identified to have mediated interaction effects on the relationship between amyloid beta accumulation and memory abilities. This identification of temporal thickness as a mediator is consistent with the existing literature (e.g., \cite{villeneuve_cortical_2014}). Further, we found that when there is no amyloid beta accumulation, a thinner cortex in these two identified regions is associated with worse cognitive performance, but when there is more amyloid beta accumulation involved, the mediation effect in these two regions disappears, which helps illustrate the brain compensation mechanism. 
One limitation is that when the number of potential mediators or the sample size becomes larger, it may take a longer time to run the algorithm, as the estimation of $\Sigma^{-1}$ for the HBIC computation becomes slower. 
In summary, our algorithm works well and can be used as an effective tool to identify the mediated interaction with preserved hierarchical structure in the mediation analysis.

\section*{Software} \label{sec6}

The XMInt R package is available at https://github.com/ruiyangli1/XMInt, and the package usage with examples is available at https://ruiyangli1.github.io/XMInt/.

\section*{Disclosure statement}

The authors report there are no competing interests to declare.

\section*{Funding}

This work was supported by NIH R01AG062578 (PI: Lee). Data collection and sharing for this project was funded by the Alzheimer's Disease Neuroimaging Initiative (ADNI) (National Institutes of Health Grant U01 AG024904) and DOD ADNI (Department of Defense award number W81XWH-12-2-0012). ADNI is funded by the National Institute on Aging, the National Institute of Biomedical Imaging and Bioengineering, and through generous contributions from the following: AbbVie, Alzheimer’s Association; Alzheimer’s Drug Discovery Foundation; Araclon Biotech; BioClinica, Inc.; Biogen; Bristol-Myers Squibb Company; CereSpir, Inc.; Cogstate; Eisai Inc.; Elan Pharmaceuticals, Inc.; Eli Lilly and Company; EuroImmun; F. Hoffmann-La Roche Ltd and its affiliated company Genentech, Inc.; Fujirebio; GE Healthcare; IXICO Ltd.; Janssen Alzheimer Immunotherapy Research \& Development, LLC.; Johnson \& Johnson Pharmaceutical Research \& Development LLC.; Lumosity; Lundbeck; Merck \& Co., Inc.; Meso Scale Diagnostics, LLC.; NeuroRx Research; Neurotrack Technologies; Novartis Pharmaceuticals Corporation; Pfizer Inc.; Piramal Imaging; Servier; Takeda Pharmaceutical Company; and Transition Therapeutics. The Canadian Institutes of Health Research is providing funds to support ADNI clinical sites in Canada. Private sector contributions are facilitated by the Foundation for the National Institutes of Health (www.fnih.org). The grantee organization is the Northern California Institute for Research and Education, and the study is coordinated by the Alzheimer’s Therapeutic Research Institute at the University of Southern California. ADNI data are disseminated by the Laboratory for NeuroImaging at the University of Southern California.

\section*{Data availability statement}

Data used in preparation of this article were obtained from the Alzheimer's Disease Neuroimaging Initiative (ADNI) database (\url{adni.loni.usc.edu}), under the data use agreement by ADNI. 
The simulation experiment data example is available at https://ruiyangli1.github.io/XMInt/articles/Simulation.html.

\newpage

\section*{Tables and Figures}

\begin{table}[h]
\caption{XMInt Algorithm Summary} \label{Table1}
\centering
\begin{tabular}{ll}
\toprule
\textbf{Input:} & $\mathbf{X}$, $\mathbf{Y}$, potential $\mathbf{M}$ \\
\textbf{Output:} & Selected mediator(s) and exposure by mediator interaction(s) \\
\midrule
\textbf{1. Initialization: } 
 & $\bullet$ standardize $\mathbf{X}$, $\mathbf{Y}$, $\mathbf{M}$ \\
 & $\bullet$ using standardized data, generate an exponentially decaying sequence \\
 & \ \ \ $\lambda_{max} = \lambda_{1} > \lambda_{2} > \cdots > \lambda_{K} = \lambda_{min}$ \\
 & \ \ \ \ \ $\circ$ compute $\lambda_{max} = n^{-1} \max \mid \mathbf{ [ X \ M \ X \times M]}^\top \mathbf{Y} \mid$ \\
 & \ \ \ \ \ $\circ$ set $\lambda_{min} = \zeta \lambda_{max}$ with some small $\zeta > 0$ \\
 & $\bullet$ $\mathcal{M}_0 = \emptyset$, $\mathcal{I}_0 = \emptyset$ \\
\textbf{2. Find regularization path: } & For each $k = 1, \cdots, K$, \\
 & $\bullet$ minimize (\ref{obj_func}) w.r.t $\boldsymbol{a,b_1,b_2}$ and get estimation \\
 & \ \ \ \ \ - \textit{regarding the penalization in (\ref{obj_func})} \\
 & \ \ \ \ \ \ \ \ \ \ $\cdot$ \textit{If $v \in \mathcal{M}_{k-1}$, then do not penalize $a_{v}$ and $b_{1_v}$} \\
 & \ \ \ \ \ \ \ \ \ \ $\cdot$ \textit{If $v \in \mathcal{I}_{k-1}$, then do not penalize $a_{v}$, $b_{1_v}$ and $b_{2_v}$} \\
 & $\bullet$ update $\mathcal{M}_k$ and $\mathcal{I}_k$ \\
 & \ \ \ \ \ $\circ$ $v$'s with nonzero $a_v$ and $b_{1_v}$ form $\mathcal{M}_k$ \\ 
 & \ \ \ \ \ $\circ$ $v$'s with nonzero $b_{2_v}$ form $\mathcal{I}_k$ \\ 
 & $\bullet$ compute HBIC \\
\textbf{3. Model selection: } & select the final model with the smallest HBIC \\
\bottomrule
\end{tabular}
\end{table}

\newpage

\begin{table}[h]
\centering
\caption{Participants' Characteristics (ADNI Dataset)} \label{Table2}
\begin{tabular}{@{}lrrr@{}}
%\begin{tabular}{@{}p{7.5cm}@{}rrr}
\toprule
\textbf{Characteristic} & \textbf{Overall}, N = 734\textsuperscript{1} & \textbf{CN}, N = 276\textsuperscript{1} & \textbf{MCI}, N = 458\textsuperscript{1} \\
\midrule 
\textbf{Abeta accumulation} (pg/mL) & 1,088.5 (447.3) & 1,229.3 (440.6) & 1,003.6 (430.1) \\ 
\textbf{Memory composite scores} & 0.6 (0.7) & 1.1 (0.6) & 0.3 (0.7) \\ 
\textbf{Age} (years) & 71.9 (7.0) & 72.8 (5.9) & 71.4 (7.6) \\ 
\textbf{Sex} &  \\ 
\ \ Male & 381 (51.9\%) & 121 (43.8\%) & 260 (56.8\%) \\ 
\ \ Female & 353 (48.1\%) & 155 (56.2\%) & 198 (43.2\%) \\ 
\textbf{Education} (years) & 16.3 (2.6) & 16.5 (2.5) & 16.2 (2.7) \\ 
\textbf{Ethnicity} &  \\ 
\ \ Hispanic & 24 (3.3\%) & 11 (4.0\%) & 13 (2.8\%) \\ 
\ \ Not Hispanic & 706 (96.2\%) & 262 (94.9\%) & 444 (96.9\%) \\ 
\ \ Unknown & 4 (0.5\%) & 3 (1.1\%) & 1 (0.2\%) \\
\textbf{Race} &  \\ 
\ \ White & 681 (92.8\%) & 252 (91.3\%) & 429 (93.7\%) \\
\ \ Non-White & 51 (6.9\%) & 24 (8.7\%) & 27 (5.9\%) \\ 
\ \ Unknown & 2 (0.3\%) & 0 (0.0\%) & 2 (0.4\%) \\
\bottomrule
\small{\textsuperscript{1}n (\%); Mean (Standard Deviation)}\\
\small{CN: cognitively normal; MCI: mild cognitive impairment}\\
\end{tabular}
\end{table}

\newpage

\newpage

\newpage

\newpage

\bibliographystyle{unsrtnat}
\bibliography{references}

\end{document}